\documentclass{aa}  
\graphicspath{{/home/quintana/SCIENCE/vycma/ALMA/Papers/1st/FIGS/}}
\usepackage{graphicx}
\usepackage{natbib}
\usepackage{txfonts}
\newcommand{\kms}{\mbox{km~s$^{-1}$}}

\newcommand{\ms}{\mbox{$M_{\odot}$}}

\newcommand{\secp}{\mbox{\rlap{.}$''$}}
\newcommand{\secs}{\mbox{\rlap{.}$^{\rm s}$}}

\newcommand{\h}{$^{\rm h}$}
\newcommand{\m}{$^{\rm m}$}

\begin{document}

   \title{History of two mass loss processes in VY\,CMa
   \thanks{This paper makes use of the following ALMA data: ADS/JAO.ALMA\#2017.1.00558.S. 
  ALMA is a partnership of ESO (representing its member states), NSF (USA) and NINS (Japan), 
  together with NRC (Canada) and NSC and ASIAA (Taiwan) and KASI (Republic of Korea), in 
  cooperation with the Republic of Chile. The Joint ALMA Observatory is operated by ESO, AUI/NRAO and NAOJ.}}

   \subtitle{Fast outflows carving older ejecta}

   \author{G. Quintana-Lacaci
          \inst{1}
          \and
          L. Velilla-Prieto
          \inst{1}
          \and
          M. Ag\'undez
          \inst{1}
          \and
          J.P. Fonfr\'ia
          \inst{2}
          \and
          J. Cernicharo
          \inst{1}
          \and
          L. Decin
          \inst{3}
          \and
          A. Castro-Carrizo
          \inst{4} 
          }

   \institute{Dept. of Molecular Astrophysics. IFF. CSIC. C/ Serrano 123, E-28006, Madrid, Spain\\
              \email{guillermo.q@csic.es}             
         \and
             Centro de Astrobiolog\'ia (CAB), CSIC-INTA. Postal address: ESAC, Camino Bajo del Castillo s/n, 28692, Villanueva de la Ca\~nada, Madrid, Spain
        \and             
             Instituut voor Sterrenkunde, KU Leuven, Celestijnenlaan 200D, 3001 Leuven, Belgium
         \and
             Institut de RadioAstronomie Millim\'etrique, 300 rue de la Piscine,
38406 Saint Martin d'H\'eres, France\\
             }

   \date{Received September 15, 1996; accepted March 16, 1997}

% \abstract{}{}{}{}{}
% 5 {} token are mandatory
 
  \abstract
   {Red supergiant stars (RSGs, $M_{init}=10-40\ms$) are known to eject large amounts of material, 
as much as half of their initial mass during this evolutionary phase. 
However, the processes powering the mass ejection in low- and intermediate-mass stars do not work for RSGs and the mechanism that drives the ejection remains unknown. 
Different mechanisms have been proposed as responsible for this mass
ejection including Alfv\'en waves, large convective cells, and magnetohydrodynamical (MHD) disturbances at the photosphere, but so far little is known 
about the actual processes taking place in these objects.}
   {Here we present high angular resolution interferometric ALMA maps of VY\,CMa continuum and molecular emission,
which resolve the structure of the ejecta with unprecedented detail.
The study of the molecular emission from the ejecta around evolved stars has been shown to be an essential tool
in determining the characteristics of the mass loss ejections. Our aim is thus to  use the information provided by these
observations to understand the ejections undergone by VY\,CMa and to determine their possible origins.}
   {We inspected the kinematics of molecular emission observed. We obtained position-velocity diagrams and reconstructed the 3D structure of the gas traced by
the different species. It allowed us to study the morphology and kinematics of the gas traced by the different species surrounding VY\,CMa.  }
   {Two types of ejecta are clearly observed: extended, irregular, and vast
ejecta surrounding the star that are carved by localized fast outflows.
The structure of the outflows is found to be particularly flat. %structure. 
We present a 3D reconstruction of these outflows and proof of the
carving. 
This indicates that two different mass loss processes take place in this massive
star.
We tentatively propose the physical cause for the formation of both types of structures.
These results provide essential information on the  mass loss processes of RSGs and
thus of their further evolution.}
   {}

   \keywords{   Stars: supergiants --
                radio lines: stars -- 
                circumstellar matter --
                stars: mass-loss --
                stars: individual: VY\,CMa
               }

   \maketitle
%
%________________________________________________________________

\section{Introduction}

Asymptotic giant branch (AGB) stars and red supergiant (RSG) stars, two categories of  evolved stars,  are  known to be among  the main dust
factories in the Universe.
Along with this dust, these objects reinject gas composed by processed material 
into the interstellar medium (ISM), 
and are thus   essential to the chemical enrichment of galaxies \citep{Hofner2018}.
While the contribution of RSG stars to the ISM dust in the Milky Way is expected
to be low ($\sim$1\%), they are thought to be the dominant source of 
ISM dust in starburst galaxies and in galaxies presenting long loop-back times (i.e., very distant galaxies) \citep{Massey08}.
The mass loss of these objects determines their fate \citep{Smith14}:  which objects reach the 
supernova phase, when   this phase takes place, and 
the contribution of heavy elements to the ISM by these stars.

The mechanism responsible for the mass ejection in low- and intermediate-mass evolved 
giant stars ($M_{init}\sim 0.6-9\ms$) during their evolution along the AGB is relatively well established \citep{Hofner2018}. Due to stellar pulsations a 
quasi-static layer is formed around the star. In this layer the temperature 
is low enough to activate dust formation. The radiation pressure on the dust and the 
dust--gas coupling \citep{dustgascouple} generate an approximately  isotropic mass ejection.

Red supergiant stars are also known to lose large amounts of mass \citep{dejager98}. However,
the pulsations of these stars are irregular and, more importantly, they have  low amplitude \citep{josselin07}, 
which prevents the formation of the quasi-static layer where the dust formation could take place.
Therefore, the AGB dust-driven mass loss mechanism does not work in the case of massive stars.
While several possibilities have been proposed, the actual mechanisms that drive mass loss ejection in 
this type of stars have not yet been established \citep{bennett2010}.

Studying the mass ejections around evolved stars has been found to be a powerful
tool to understand the mass loss processes. In particular, 
the molecular emission from these ejecta provides spatio-kinematical information
which allows us to reconstruct the events generating the ejection. These results provide 
strong constraints on the mechanisms driving those events.
 
Large convective cells have been identified in the closest RSG star, Betelgeuse \citep{Lim98},
and have been associated with a
trefoil structure observed around the photosphere of the star with the ALMA interferometer \citep{kervella18}. 
However, despite Betelgeuse's relatively close distance \citep[197 pc;][]{Harper08}, the structure 
of the inner ejecta has not been resolved enough to disentangle
between the different mass ejection mechanisms proposed. 

On the other hand, VY\,CMa, a RSG located at a distance of 1.2\,kpc \citep{Zhang12}, presents a massive  
ejecta very rich in molecular species with an oxygen-rich chemistry \citep{Zyuris07, Tenenbaum10}
that can be studied in detail.
The ejecta around this object is known to present a complex structure
\citep[e.g.,\ as seen in Hubble Space Telescope images][]{kastner98,Humphreys07}.
Accordingly, the molecular lines show complicated line profiles, which
suggest a heterogeneous spatio-kinematical structure. Thanks to the use of the ALMA interferometer, 
in particular to the molecular maps obtained by  \citet{debeck15}, \citet{Decin16}, and 
\citet{ogorman15}, among others,  the complexity observed in the outer regions is also observed in the innermost areas of the ejecta.

\citet{Richards14} identified two main regions (see Fig.\,1): $VY$, which 
corresponds to the central star and presents a N-E elongation, and $C$, which is associated 
with very cold and dense dust %(450\,K, $\sim 2 \times 10^{-4}\ms$)\cite{OGorman}. 
($T_d \sim 100\,K$,  $\sim 1.2 \times 10^{-3}\ms$) \citep{Vlemmings17}.
The maps obtained by \citet{debeck15} for the TiO$_2$ emission and \citet{Decin16} 
for the NaCl emission present the same spatial features, for a spatial 
resolution of $\sim$0\secp1--0\secp2.
These regions have also been identified in our maps with higher spatial resolution (Fig.\,1). 
Other regions identified in previous works are also shown in Fig.\,1.

\citet{ogorman15} estimated that the mass loss events responsible 
for the formation of the different regions surrounding VY took 30--50\,years. 
These authors suggest  that on these timescales  the formation of these structures 
cannot be a consequence of the presence of convective cells; although the direction 
of the ejections took place at different angles, the duration of the mass ejections 
was relatively long compared to typical timescales of the convective cells 
($\sim 150$ days). They suggested that the formation of long-lived cold spots 
at the stellar atmosphere, related with magnetic fields, where dust can be 
formed and drive a somewhat collimated mass ejection, is a better  
mechanism responsible for the formation of the structures observed.
It is worth noting that these cold-spot induced ejections would present a  
velocity field similar to the ordinary dust-driven mass ejections, reaching a terminal 
velocity related with the dust properties and the luminosity of the star.

The origin of the structures observed toward VY\,CMa has also been associated with 
a dimming period undergone by VY\,CMa $\sim$ 100 years ago \citep{Kaminski19}. A similar 
process seems to be taking place now in Betelgeuse. In  that sense Betelgeuse 
might be an object at an earlier stage than VY CMa. Recent works \citep{levesque20} 
suggest that the decrease in its brightness is related with the formation of dust 
in our direction. However, an intensity decrease has also been observed at millimeter-wavelengths
which might suggest that the changes are not due to the formation of a localized mass 
ejection in our direction, but to changes in the photosphere itself \citep{Dharmawardena_2020}.
More recently it has been suggested that both effects might be   responsible for the 
dimming observed in this object, a local decrease in the photospheric temperature 
that triggered the dust formation and ejection of material in our direction \citep{montarges21}. 

In order to study the characteristics of the ejecta, and the processes driving these mass ejections
we obtained ALMA molecular maps of VY\,CMa with an angular resolution of $\sim 25-35$\,mas 
at the spectral range $\sim$216.20 -- 235.45\,GHz which covers molecular lines from species 
such as NaCl, H$_2$S, H$_2$O, SiO, SiS, $^{13}$CO, and PN, among others (Table 1). The 
obtained maps provide unprecedented detail of the structure of the ejecta that allow us to 
propose for the first time a 3D view of the ejecta. The description of the observations 
and data reduction is presented in   Sect.\,\ref{Obs} of the present work.

\section{Observations}
\label{Obs}
The observations of VY\,CMa were performed with the ALMA interferometer
(project code 2017.1.00558.S, in Cycle 5) during a unique observing
execution of 1.3h in October 2017.  Forty minutes of acquisitions were
obtained on the source, with 51 antennas. Baselines ranged from 41m to
16km, providing an angular resolution $\sim 0\secp025$ and with maximum
recoverable scale for the most extended emission of $0\secp36$. The
amount of precipitable water vapor was $\sim\,1$mm, humidity 10-35\%,
ambient temperatures $\sim -4^{\circ}$, and system temperatures 60--80 K.
Four spectral windows of 1.875 GHz bandwidth were centered at
the sky frequencies 217.090, 219.974, 232.641, and 234.516 GHz, with
spectral resolutions of 1.35, 1.33, 1.26, and 1.25 km/s.

The bright point-like source J0522-3627 was observed to calibrate the
bandpass, and was also set as absolute flux reference, with fluxes of
4.17, 4.16, 4.10, and 4.09 Jy for each spectral window.  Observations
of J0725-2640 were performed every two minutes to calibrate amplitude
and phase gains in time. The phase rms after calibration was 32 deg.

The standard phase calibration was subsequently improved by
self-calibration on the continuum emission from the
source. Imaging restoration was made with a robust
weighting.  
Image synthesis was made
by using the cleaning method SDI\citep{SDI}. The obtained synthetic beam is 29 $\times$ 25 mas, and
the brightness rms per channel 
is $\sim$ 0.15 mJy Beam$^{-1}$.
The brightness rms achieved in the continuum map is in the range 0.028--0.052 mJy Beam$^{-1}$ 
depending on the spectral window (spw), with a continuum emission peak at R.A.  
07\h22\m58\secs323 Dec. -25$^\circ$46${'}$03\secp00
(J2000). 

Continuum was also subtracted by averaging the visibility channels free of molecular emission
for each spectral window,  
and subtracting these continuum visibilities
from the original data. 
Calibration was performed with the CASA\footnote{https://casa.nrao.edu/} software package, and data
analysis with GILDAS\footnote{https://www.iram.fr/IRAMFR/GILDAS/} and {\tt Astructures\footnote{https://github.com/GQuintanaL/Astructures}}.

\begin{figure}
   \centering
   \includegraphics[angle=0,width=8cm]{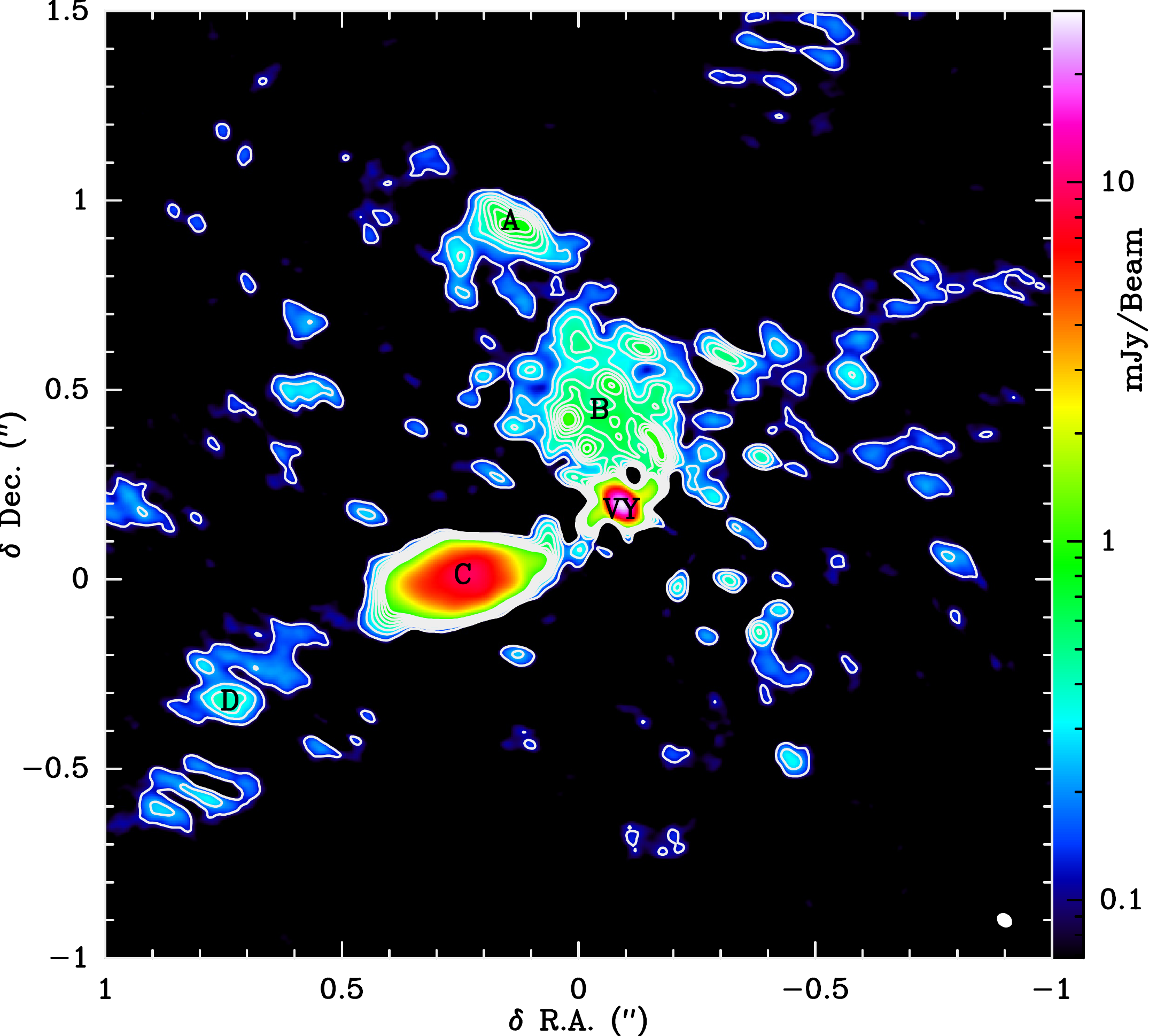}%
\caption{Continuum emission obtained from the ALMA data at $\lambda=\,1\,mm$. 
The half power beam-width is $0\secp04\times0\secp03$ with a position angle 
 of 29$^{\circ}$ (see bottom right corner). The contours correspond to steps of 
5$\sigma$ ($\sigma$ = 0.023 mJy/beam). The labels of the different regions follow 
\citet{Kaminski19}. 
\label{cont}
}
\end{figure}

\section{Structure of the ejecta}

\begin{figure*}[tbh!]
        \centering
        \includegraphics[scale=0.7]{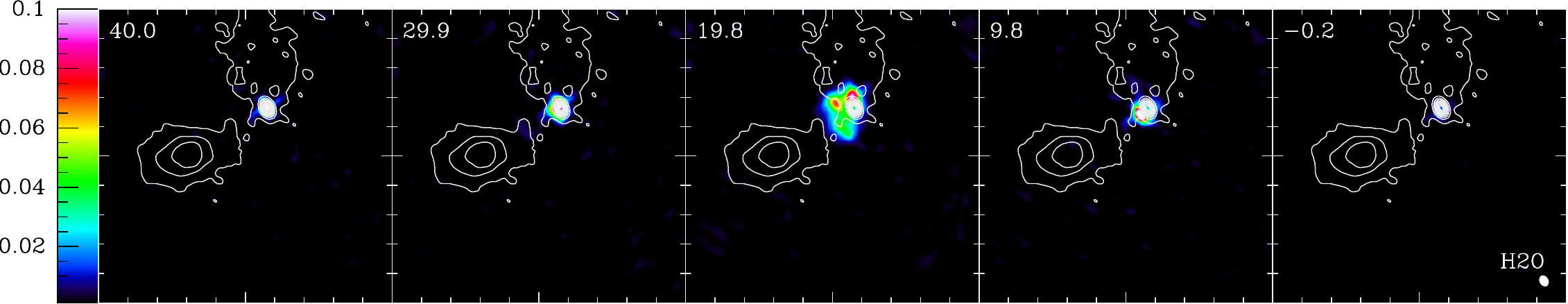}
        \vspace{-1mm}
        \hspace*{-0.5mm}\includegraphics[scale=0.7]{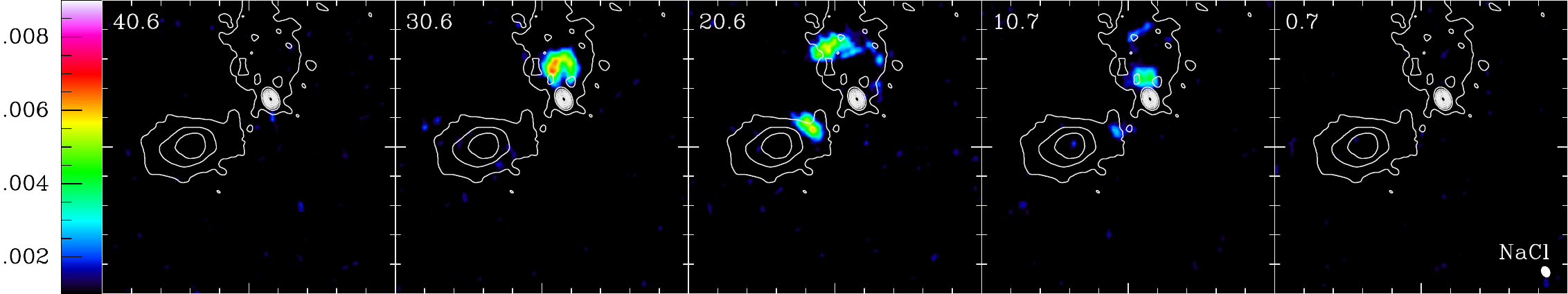}
        \vspace{-1mm}
        \hspace*{-0.5mm}\includegraphics[scale=0.7]{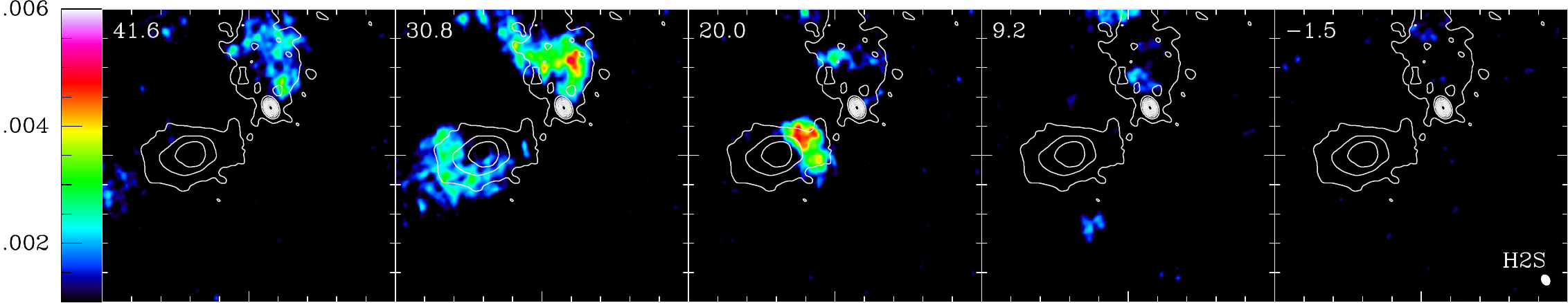}
        \vspace{-2mm}
        \includegraphics[scale=0.7]{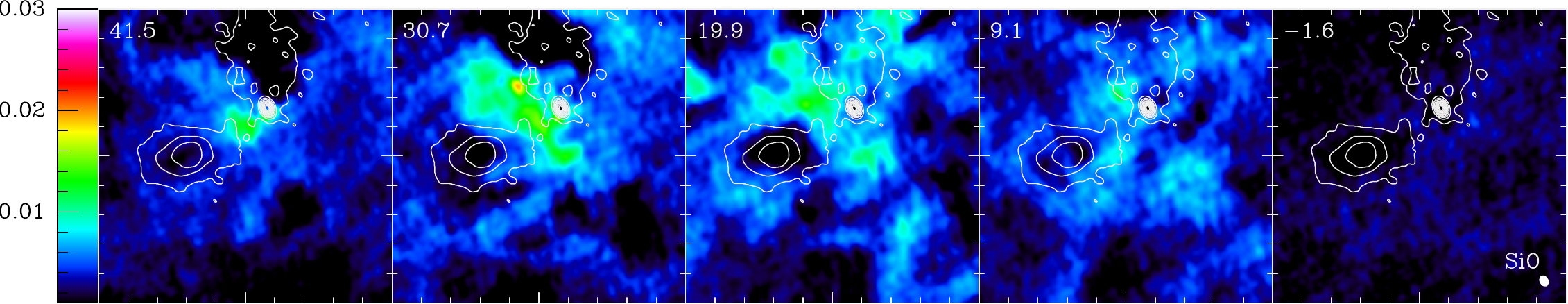}
        \vspace{-1mm}
        \hspace*{7.4mm}\vspace{0mm}\includegraphics[scale=0.7]{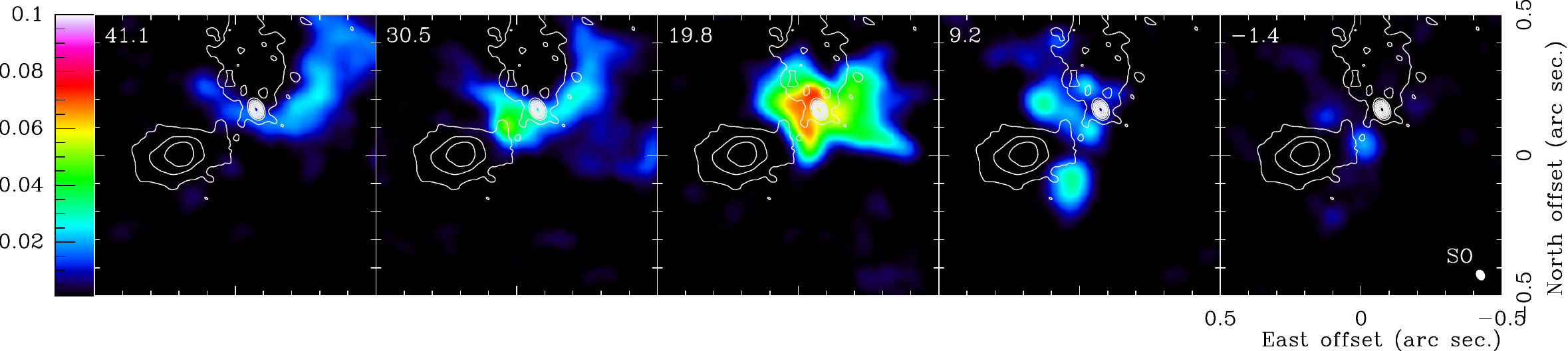}
\caption{Channel maps of some lines showing the 
complexity of VY\,CMa ejecta (continuum shown in contours). From top to bottom in color: 
H$_2$O $v=2$ $5_{5,0}-6_{4,3}$;
NaCl 18--17;
H$_2$S 2$_{2,0}$--2$_{1,1}$
SiO 5--4; 
SO $5_6-4_5$. 
The value of $V_{LSR}$ of each panel is shown in the upper left corner.
The HPBW is $0\secp04\times0\secp03$ with a P.A. of 29$^{\circ}$ (see the bottom right corner of the last
panel). The continuum emission is shown in white contours.
}
\label{maps}
\end{figure*}

The ALMA data covered a large variety of species (Table 1). The different species were identified 
using MADEX \citep{MADEX}. Some of these lines were detected, but are too weak to obtain a usable map.
Our new ALMA maps reveal that the molecular and continuum emission can be divided 
into two general types. The continuum emission (Fig.\,\ref{cont}) presents relatively 
compact structures already identified in previous works \citep{Richards14,ogorman15,debeck15}. 
These structures are also clearly traced by certain species such as H$_2$S or NaCl. 
On the other hand, the emission from SiO, SiS, SO or $^{13}$CO presents a more extended 
emission. See Fig.\,\ref{maps} for a general reference of these structures.  
The extended emission is partially filtered out. However, it can be clearly seen that 
these large-scale  emission data present hollow regions.

The emission of SiO and other species such as SO or SO$_2$ is absent in those 
regions where H$_2$S emission arises (see Fig.\,\ref{maps} and Fig.\,\ref{SO_H2S} 
for a full SO channel map). In particular, SO presents a series of bubble-like 
structures in its ejecta. 
Thanks to the detailed radiative transfer modeling of the line emission of SO 
and SO$_2$ in this object, it has been noted that the origin of 
these species might be inner material swept up by high-velocity winds\citep{Adande13}. While 
SiO is more ubiquitously distributed, SO would be a better tracer of the 
edge of these cavities.

As mentioned by \citet{ogorman15}, this extended ejecta 
seems 
to correspond to a quiescent, less localized mass ejection, that  has been carved 
by recent collimated ejections. SiO and SO emission present clear signs of this carving, 
as the maps present no emission of these lines in the regions where the narrow outflows arise 
(see Fig\,\ref{maps}).

In particular, we can identify in the first channels of Figs.\,\ref{maps} and \ref{SO_H2S} the SO emission 
surrounding the northern fast outflow, which is almost in the plane of the sky, but slightly tilted 
toward us. In the middle channels the SO emission surrounds clump $C$. Finally, the last channels show
a shell that coincides with the rear outflow detected in the 3D reconstruction presented below. 
This clearly suggests that these localized jets have carved gas ejected previously. These holes 
formed in the quiescent ejecta by the collimated jets might be those inferred in previous works 
\citep{Humphreys19,Humphreys21}. 
%Furthermore, as already shown\cite{Gordon} H$_2$S has been also observed in the SW clump observed in HST images, which could be related with 
%, which would probably explained the curved structure observed in the southern inner SO emission as a consequence of a prior fast and narrow ejection. Probably gas is already too diluted for H$_2$S to be detected but in the southern blob.

In addition, we find SiO and SO emission in the inner regions, closer to the central star. This 
indicates that the quiescent mass ejection has taken place both before and after the occurrence 
of the fast localized outflows.

\begin{table}
 \renewcommand{\arraystretch}{0.5}
 \begin{center}
 \caption{Parameters of the observed lines.}
 \begin{tabular}{l l l r}
 \hline\hline
  $\nu_\mathrm{rest}$ (MHz) & Molecule & Transition & $E_\mathrm{up}$ (K) \\
  \hline
%\tablenum{4}
216531.44180  & Na$^{37}$Cl  & 16--17 & 93.6\\
216643.30424  & SO$_2$ & 22$_{2,20}$--22$_{1,21}$ & 248.5\\
216710.44359  & p-H$_2$S & 2$_{2,0}$--2$_{1,1}$ & 84.0\\
216757.60348  & SiS & 12--11, v=1 & 1138.8\\
217104.91983  & SiO & 5--4 & 31.3 \\
217817.65596  & SiS & 12--11 & 68.0\\
219376.92           & U & -- & --\\
219465.54552  & SO$_2$ & 22$_{2,20}$--22$_{1,21}$, v=2 & 993.7\\
219614.93595  & NaCl & 17-16 v=1 & 614.5\\
219776.66           & U & -- & --\\
219949.39633  & SO & 5$_6$--4$_5$ & 35.0\\
220165.25185  & SO$_2$ & 16$_{3,13}$--16$_{2,14}$, v=2 & 893.4 \\
220330.80523  & AlOH & 7--6 & 42.3\\
220398.68354  & $^{13}$CO & 2--1 & 15.9\\
%221260.14696  &        NaCl & 17--16 & 95.6\\
%221965.22112  &        SO$_2$ & 11$_{1,11}$--10$_{0,10}$ & 60.4\\
%225153.70462  &        SO$_2$ & 13$_{2,12}$--13$_{1,13}$ & 93.0\\
%226300.02778  &        SO$_2$ & 14$_{3,11}$--14$_{2,12}$ & 119.0\\
%226287.41900  &        CN &\\
%226616.57100  &        CN &\\
%226887.42000  &        CN &\\
231963.56981    & TiO$_2$ & 19$_{2,18}$--19$_{1,19}$ & 137.6\\
%231983.06     &        ???EsAlgo &\\
%232618.50730  &        PH3?? &\\
232628.58461  & Si$^{33}$S & 13--12 & 78.2\\
232686.70000  & o-H$_2$O & 5$_{5,0}$--6$_{4,3}$, v=2 & 3427.8\\
232031.09           & U &-- & --\\
232509.97689  & NaCl & 18--17 v=1 & 625.7\\
233116.41893  & AlCl & 17-16, & 95.1\\
233664.46372  & SiS & 13--12 v=2 & 1150.1\\
234187.05727  & SO$_2$ & 28$_{3,25}$--28$_{2,26}$ & 403.1\\
234421.58220  & SO$_2$ & 16$_{6,10}$--17$_{5,13}$ & 213.3\\
%234426.87          & ?comp.map &\\
234251.91247  & NaCl & 18--17 & 106.9\\ 
234812.96839  & SiS & 13--12 v=1 & 1150.1\\
234935.69126  & PN & 5--4 & 33.8\\
235130.60           & U & -- & --\\
235151.72126  & SO$_2$ & 4$_{2,2}$--3$_{1,3}$ & 19.0\\
235174.27           & U &-- & --\\
\hline
\end{tabular}
\end{center}
\end{table}

\subsection{$VY-A$ \& $VY-C$ ejecta}

\begin{figure*}[th!]
   \centering
   \includegraphics[angle=0,width=8cm]{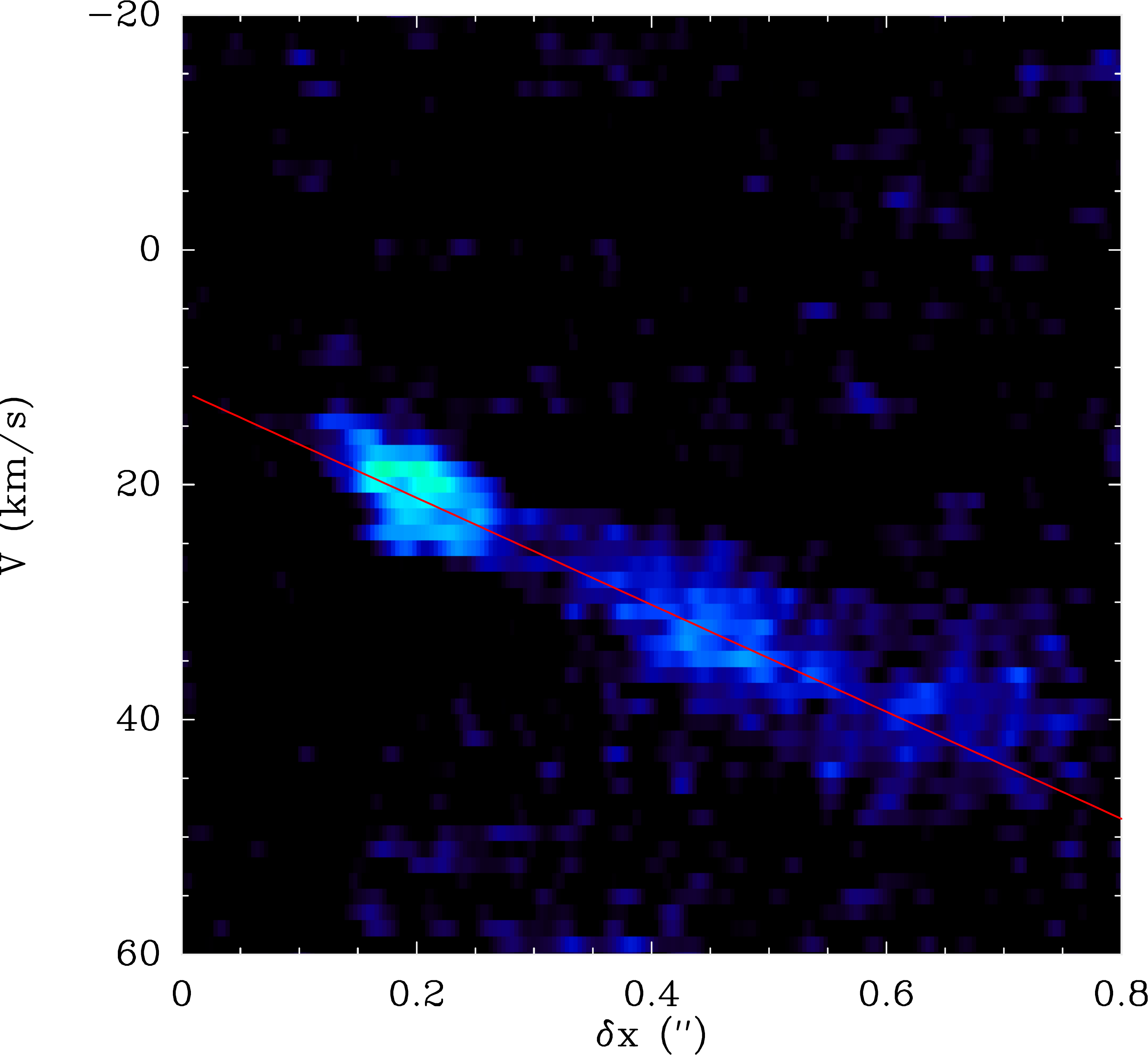}%NaCN_stack_azave.pdf} 
   \includegraphics[angle=0,width=8cm]{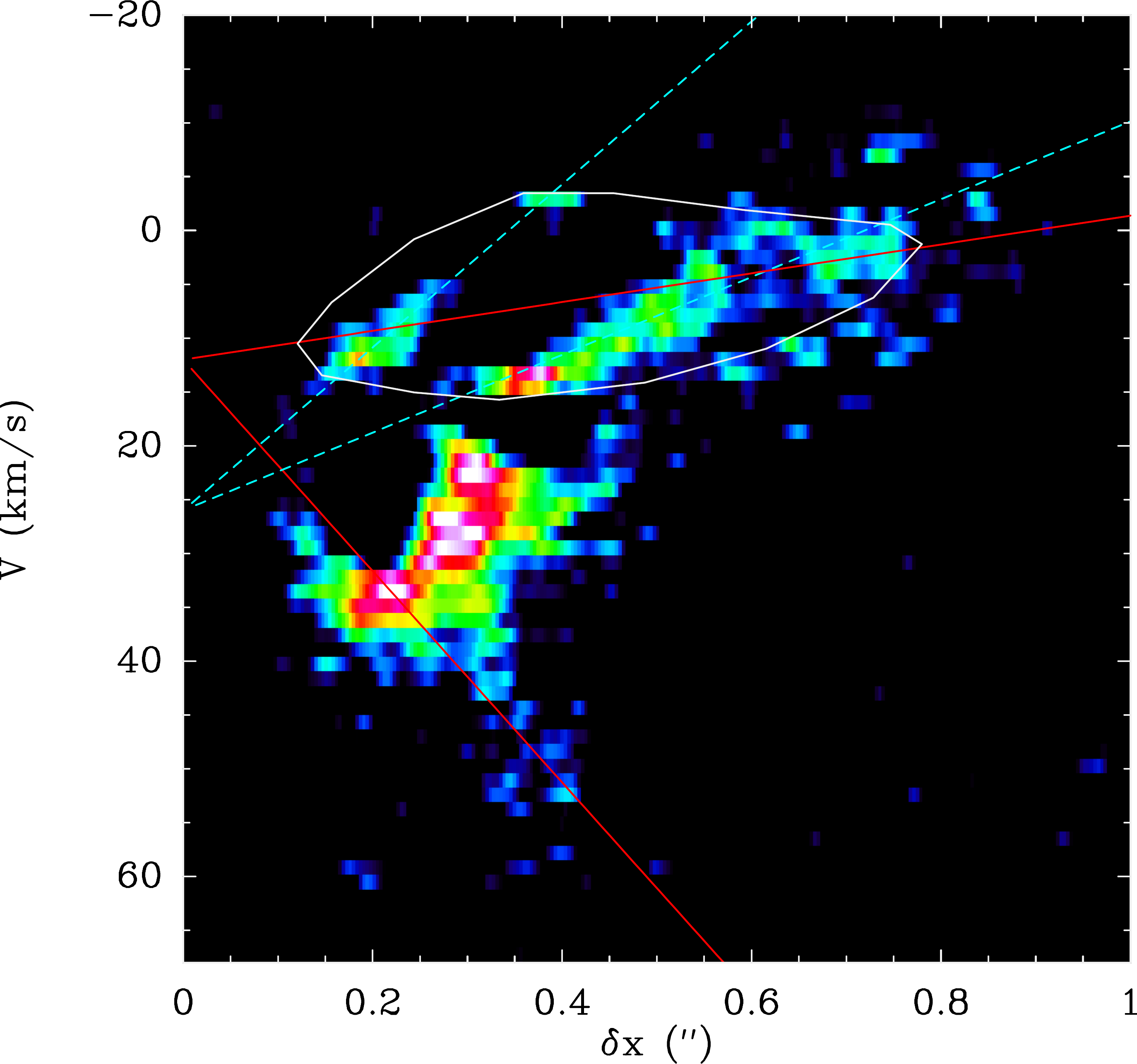}
   \caption{{Position-velocity diagrams obtained for H$_2$S 2$_{2,0}$--2$_{1,1}$ along different directions.}
   \emph{Left:} PV obtained for H$_2$S 2$_{2,0}$--2$_{1,1}$ emission along the line connecting $VY$ and $C$. The red line 
corresponds to a linear jet with an inclination of 15$^\circ$ with respect to the plane of the sky and 
$v = 17 \times r/0\secp1\, \kms$ and $V_\mathrm{LRS}=12\kms$. \emph{Right:} PV obtained for the same 
transition along the line connecting $VY$ and $A$. The
red lines correspond to the same velocity field as in the  left panel and an inclination with respect to 
the plane of the sky of 4.5$^\circ$ for the upper jet and 30$^\circ$ for the lower jet (see red lines). The white line surrounds
the suggested structure of the northern jet (see text). The cyan line corresponds to  $V_\mathrm{LRS}=26\kms$ (see text).
 }
              \label{PV1}%
\end{figure*}

As mentioned above, we observed that the interferometric maps of NaCl $v=0,1$ and H$_2$S 
traced the collimated structures $A$ and $C$ observed in previous works \citep{debeck15,Decin16}, which are also traced by
the continuum (Fig.\,\ref{cont}).
We thus used the maps of these species to study the structure of the localized mass ejections. 
We obtained position--velocity (PV) diagrams of these line emissions along the 
main axis of the two structures connecting them with the central star $VY$. In 
the case of NaCl and H$_2$S these diagrams showed that these outflows
present a straight-line shape; in other words, they follow a Hubble-like expansion velocity field ($V_{exp} \propto r$; see Fig.\,\ref{PV1}). 
In these PV maps we can see that while the ejecta $\emph{C}$ consists of a relatively simple 
outflow, the structure of the $\emph{A}$ outflow is  more complex, probably 
as this latter structure is composed   of different outflows. 
We note that the region $B$ is included in the $VY-A$ outflow, 
and we do not refer to it explicitly in the rest of the text.
In order to fit the velocity field of these outflows, we took into account a twofold degeneracy. 
There is a series of pairs of values of the jet inclination with respect to the plane of the sky ($\theta$)
and the value of the velocity constant ($V_0$), which would fit the velocity field of 
the outflow.

From the literature we know that the  \emph{C} outflow is close to the plane of the 
sky \citep{Richards14,debeck15,Kaminski19}. While the 3D model presented by \citet{Kaminski19}
suggests that \emph{C} is directly in the plane of the sky, spectral maps suggest that it is slightly
tilted. We thus can assume that the inclination of \emph{C} is equal to or lower than that of blob \emph{D},
which was $\sim 15^\circ$ \citep{Kaminski19}. %In our work we will adopt a tilt of 15$^\circ$.  

Assuming this inclination for the \emph{C} jet, we found we could fit the expansion velocity to $v = 17 \times 
r/0\secp1\, \kms$. 
Its important to note that this type of velocity field ($v\propto r$) cannot be 
created by the radiation pressure on dust grains, but a different mechanism might 
be the responsible for these ejections. This type of velocity field has been 
extensively observed in the post-AGB fast outflows.
In addition, the fitting of the outflow connecting $VY$ and $C$ suggests 
the systemic velocity at $VY$ might not be 22\,\kms\ as previously thought \citep{Decin16,Richards14}, but 
$V_{LSR}=$12\,\kms. This is compatible with the asymmetries observed by de \citet{debeck15} 
for the TiO$_2$ line profiles. Obtaining the value of $V_{LSR}$ as the mean velocity 
of the profile is useful for sources with an isotropic mass loss ejection, such as AGB stars,
or for  objects with symmetric mass ejections, such as some pre-planetary nebulae (pPNe). However, for cases like VY\,CMa, 
where the mass ejections seem to take place in different directions, this procedure might provide 
incorrect results.

\begin{figure*}
   \centering
   \includegraphics[angle=0,width=18cm]{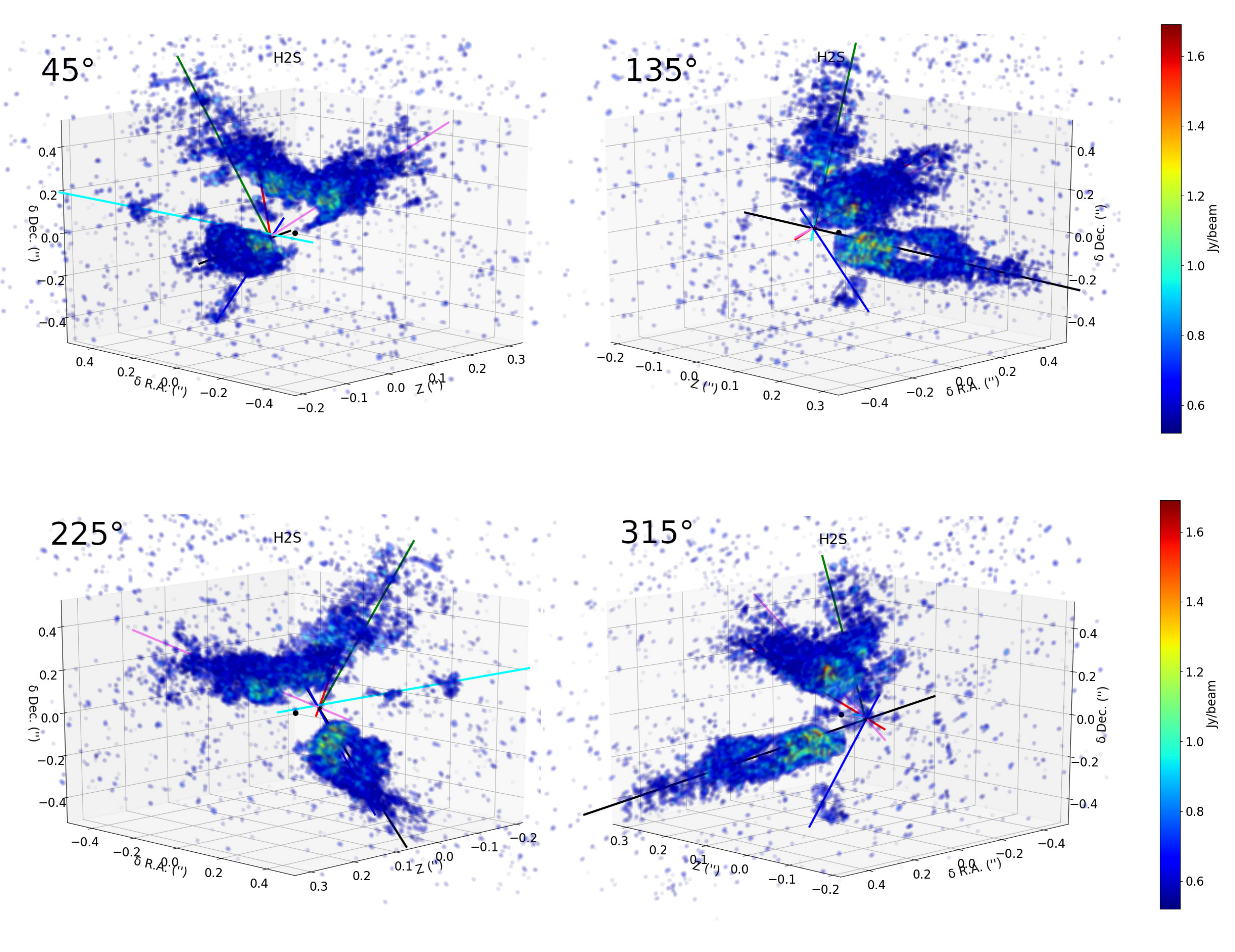}%
\caption{Reconstruction of the H$_2$S $2_{2,0} - 2_{1,1}$ emission structure.
The structure is generated for a flux with a S/N $\geq$3.
The colored arrows correspond to the main axis of the different outflows, as shown in the videos.
For more detail, see the video presenting the rotation of the H$_2$S structure.
The angle is taken in the clockwise direction with respect to the line of sight.}
\label{H2S-3d}
\end{figure*}

The PV diagram obtained along the line connecting $VY$ and $A$ presents two main jets 
(Fig.\,\ref{PV1}-right), 
one jet pointing backward and a second larger jet closer to the plane of the sky and pointing toward the 
north. The $B$ region is  composed of the emission of these two jets, while $A$    arises just 
from the latter.
The jet pointing away from us suggests a source velocity 
similar to that of jet \emph{C}, while the larger jet suggests a LSR velocity of 26\,\kms. However,
a careful inspection of the $VY$--$A$ ejecta suggests that we might not be   detecting certain 
parts of this ejection, and that the different emitting regions that are within the area delimited 
by the white line in Fig.\ref{PV1}, $right$,    correspond to the same unique ejection. 
This is better seen in the 3D reconstruction (see Fig.\,\ref{H2S-3d}). 
Nevertheless, the 
possibility of the $VY$--$A$ jet arising from $VY$ at 26\,\kms\ cannot be completely ruled out, %discarded 
and future observations will shed light on this fact.

If we impose the same velocity field, the inclination for the northern jet is  
4.5$^\circ$ and that of the rear jet is  -30$^\circ$. {It is important to note that these 
jets accurately correspond to the hollows observed in the  SO emission.}

The value of $V_\mathrm{LSR}$ obtained corresponds to a position in the 3D structure 
of the ejecta. Tracing back the trajectory of the different outflows can be used to 
confirm the real position of the central star (see Fig.\,\ref{H2S-3d}). 
 
The extent of the $VY$--$C$ and $VY$--$A$ outflows is similar, around $\sim 0\secp7-0\secp8$, which 
corresponds to $ 1.3-1.5 \times 10^{16}$cm, and reaches a velocity of $\sim 120-140\kms$ 
at the tip of these outflows. The formation of these structures took $\sim 50-60$ years, 
confirming the timescales suggested above \citep{ogorman15}. It is worth noting 
that these high velocities are not observed in the line profiles due to the 
projection effects, which means that  the ejection that dominates the width of the 
line profiles is that corresponding to the less localized  and slower mass ejection.

%%%%%%%%%%%%%%%%%%%PV DIAGRAMS

\begin{figure*}
   \centering
   \includegraphics[angle=0,width=18cm]{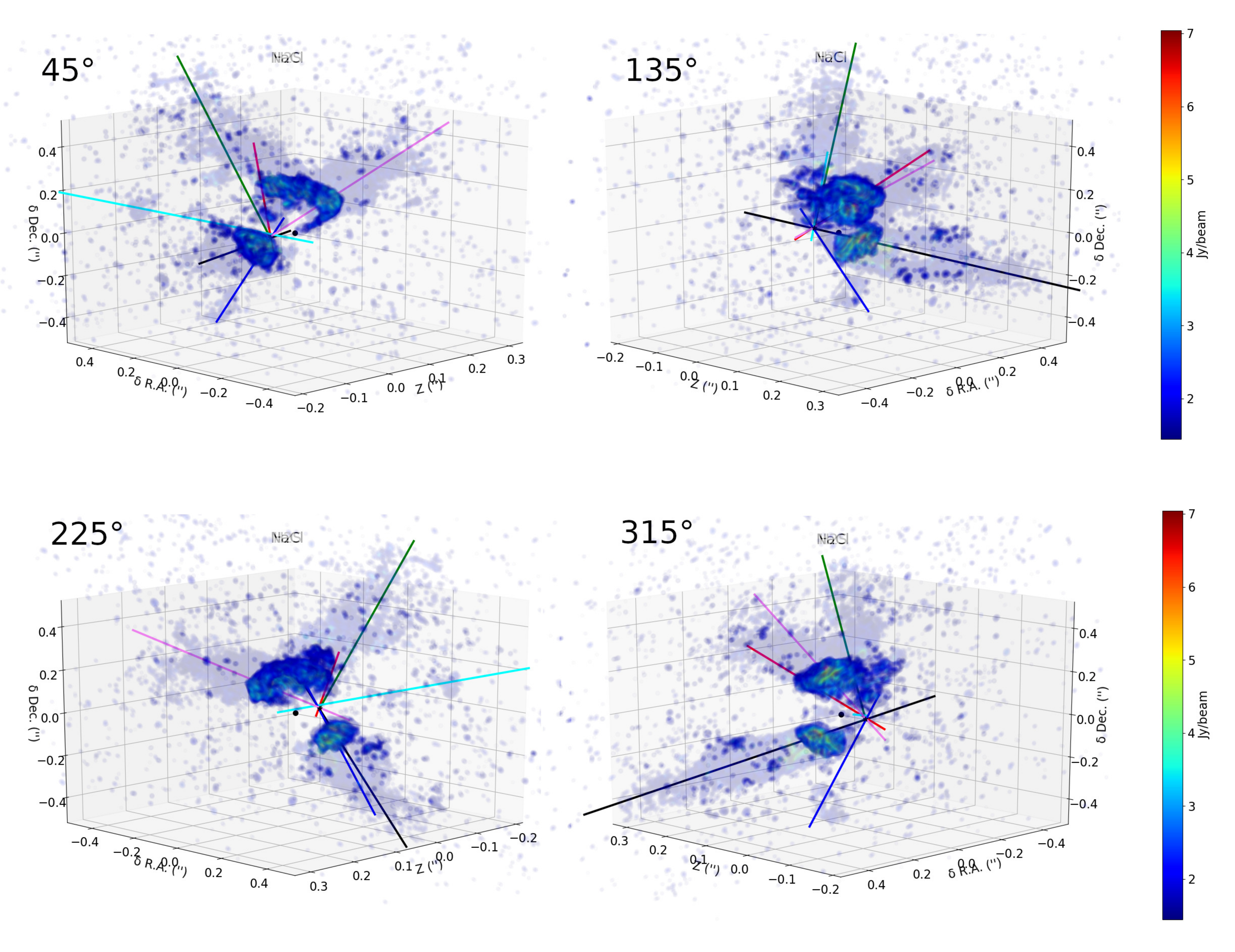}%
\caption{Reconstruction of the NaCl $J$=18--17 emission structure (colors) 
overplotted on the H$_2$S emission presented in Fig.\,\ref{H2S-3d} (transparent colors). The structure is
generated for the flux with a S/N$\geq$3. }
\label{NaCl-3d}
\end{figure*}

\begin{figure}[th!]
   \centering
   \includegraphics[angle=0,width=8cm]{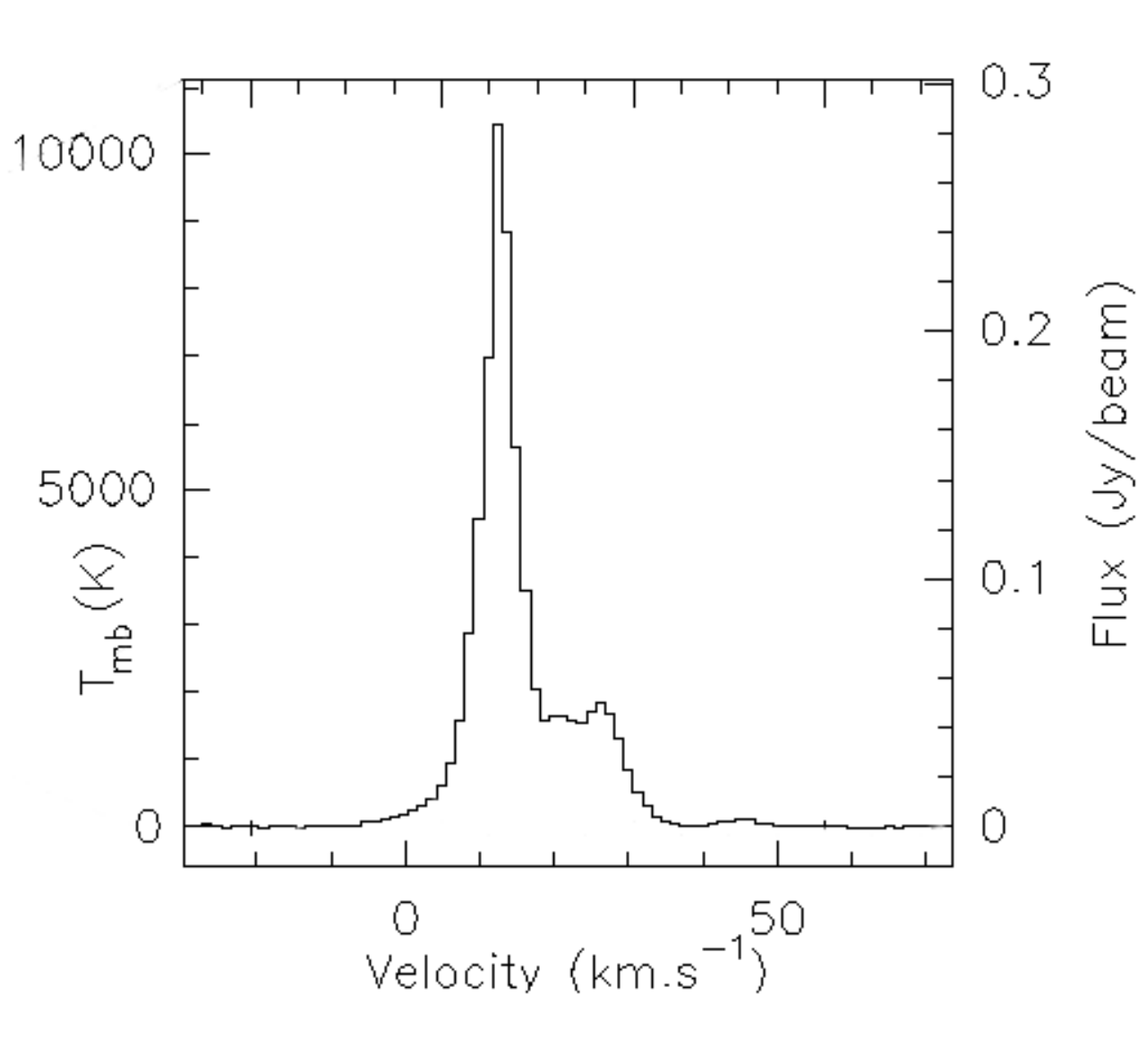}%NaCN_stack_azave.pdf} 

   \caption{Profile of the ortho-H$_2$O $5_{5,0} - 6_{4,3}, v=2$ maser emission structure toward 
   the maximum of the emission.}
              \label{h2o}%
\end{figure}

\subsection{Three-dimensional reconstruction}

Due to the complex structure of the ejecta we generated 3D structures imposing the expansion 
velocity field obtained from the PV diagrams using the {\tt Astructures 
code}. %\footnote{https://github.com/GQuintanaL/Astructures}}. 
Briefly, {\tt Astructures} reads the data from a spectral cube, estimates the
Gaussian noise from the cube, and crops out the data below a given threshold  (i.e., 3$\sigma)$.
The spectral maps consist of two spatial coordinates (R.A. and Dec.) and a velocity coordinate. 
The code plots the points in a flux scale following these coordinates.
Assuming a given velocity field, this dimension can be transformed into a space dimension, 
allowing us to reconstruct a 3D spatial structure. In addition, to help us identify the different
structures, the code rotates the structure around a selected axis, creating images for each 
rotation angle and a video.

The  reconstructed structure is rotated to allow the inspection of the different elements. 
These 3D structures rotated are presented in Figs.\,\ref{H2S-3d} and \ref{NaCl-3d} and 
in the videos associated with this work. 

Thanks to the 3D reconstruction, at least six outflows can be identified in the structure 
obtained, especially from the H$_2$S maps (see red arrows in Fig.\,\ref{H2S-3d}). Two outflows correspond to 
the structures observed in the continuum and molecular emission \citep{debeck15,Decin16,ogorman15} 
(outflows $VY$--$C$ and $VY$--$A$), while we identified two other outflows placed 
behind the $VY$--$A$ outflow. For the last two outflows only bullet-like structures are visible. We cannot determine
if the bullet-like structure of these outflows is real or due to a lack of S/N in the obtained maps.

On the contrary, the PV diagrams obtained from species like SiO or SO present very 
complex structures, probably as a consequence of swept-up and shocked regions \citep{Adande13}.
The flux filtering affects structures with scales larger than $\sim$0\secp4. 
However, this only affects homogeneous structures. 
Inhomogeneous structures are preserved. Furthermore, under certain circumstances filtering 
the homogeneous large-scale structures allows a better tracing of the mid-size and small 
  structures, due to the relative dynamic range.

\begin{figure*}[th!]
   \centering
   \includegraphics[angle=0,width=18cm]{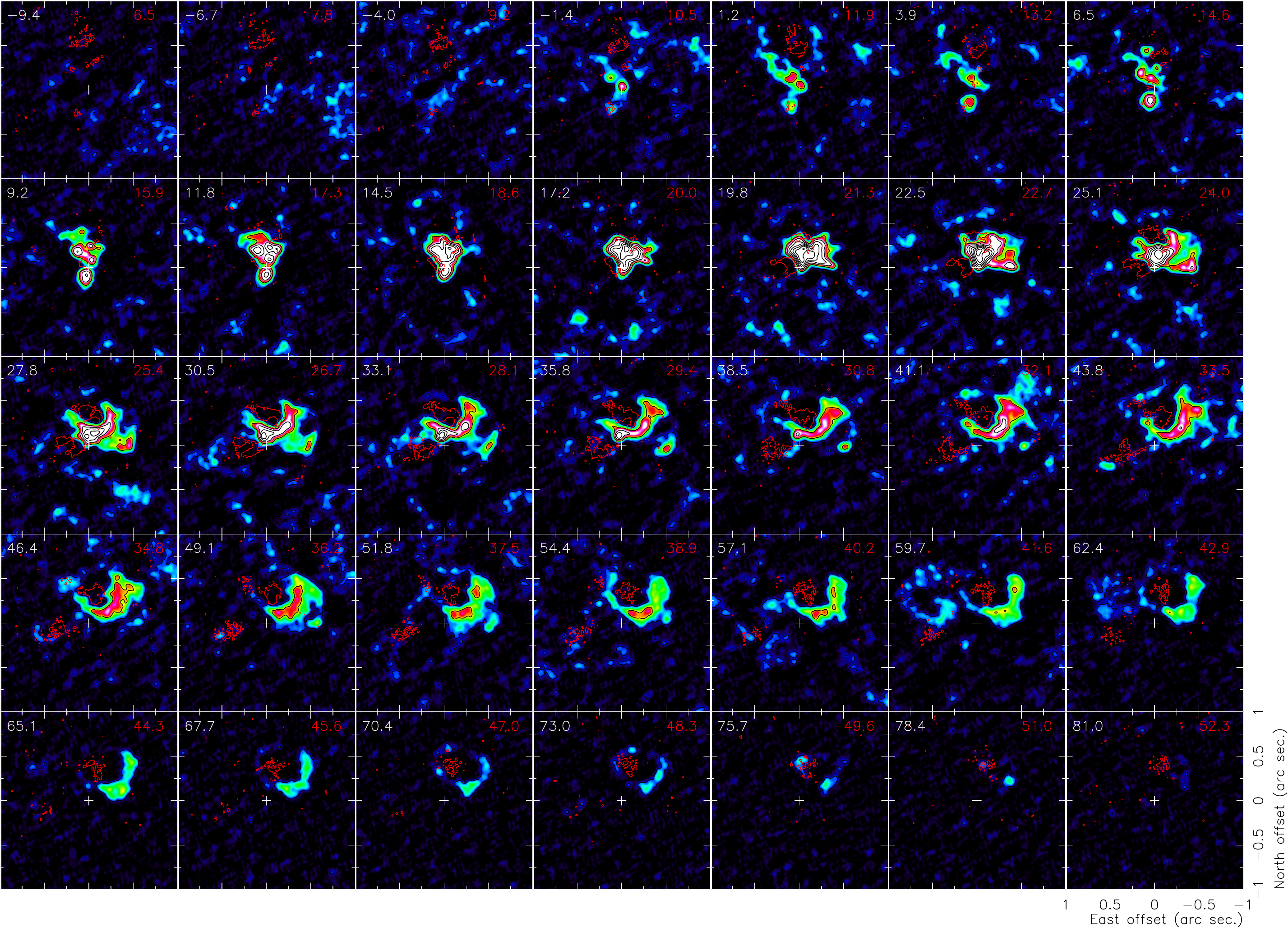}%NaCN_stack_azave.pdf} 
   \caption{Comparison of the SO $5_6-4_5$  emission (colors and black contours) and H$_2$S 2$_{2,0}$--2$_{1,1}$(red contours).
   $V_\mathrm{LSR}$ for SO is shown in the upper left corner, while that 
   of H$_2$S if shown in red in the upper right corner. A video is available as supplementary material.}
              \label{SO_H2S}%
\end{figure*}

The fast outflows observed are compatible with the bubble-like structures observed 
in SO (Fig.\,\ref{SO_H2S}). In particular, in this figure we can see 
these shells 
as a consequence of the carving by the jets presented above, at 
$\sim 1.2\kms$, $\sim 30.5\kms$, and $\sim 59.7\kms$, and as a shell extending along the channels in 
the range $\sim 30-80 \kms$. 
As mentioned, the velocity field of SO is particularly complex, and different from that 
observed for the fast outflows. This prevented a 3D reconstruction and 
a direct comparison of the structures traced by SO and H$_2$S.
In order to inspect the relation of the fast outflows with the bubbles 
observed in SO, we plotted the H$_2$S emission (overplotted   the SO 
channel maps), for which we modified the velocities of H$_2$S so the 
shells and the outflows coincide (see Fig.\,\ref{SO_H2S}, where the 
velocity of both emissions are shown).
This comparison shows that 
the outflows traced by H$_2$S and the shells probed by SO are related. 
In particular it suggests that the outflows are carving previous ejecta 
and forming cavities. 
{This coincidence of the jets and the holes clearly suggests the 
existence of two mass loss processes, and the carving of previously 
ejected material by the fast outflows.}

\subsection{Inner ejecta}

We also obtained inteferometric maps of the maser emission of ortho-H$_2$O $5_{5,0}-6_{4,3}$, ${v_2=1}$ at 232\,GHz. 
This emission presents a local intense peak at $\sim 12\kms$, which is located at R.A. 
07\h22\m58\secs327 Dec.  -25$^\circ$46${'}$03\secp06 (J2000).  This maser was predicted 
to be located much closer to the star than the  other water masers \citep{Menten89}. Since 
the pumping mechanism of the maser is the IR emission from the central star, and the velocity 
of the peak corresponds to what we found to be compatible with the origin of the
fast outflows, we propose that the position of this peak 
might correspond with the real position of the central star. Furthermore, the peak position 
of the continuum is not compatible with the expected origin of the H$_2$S outflows.  
In a similar way to $V_{LSR}$, the general assumption that the peak of the continuum emission corresponds
to the stellar position is valid
in those cases where the ejection of dust is isotropic or presents some type of symmetry. 
The intense emission of molecular lines arising at the stellar photosphere, like that 
of the o-H$_2$O maser, could provide a more accurate determination of the position of 
the central star in those cases where the ejection of material is more chaotic.

\begin{figure*}[th!]
   \centering
   \includegraphics[angle=0,width=18cm]{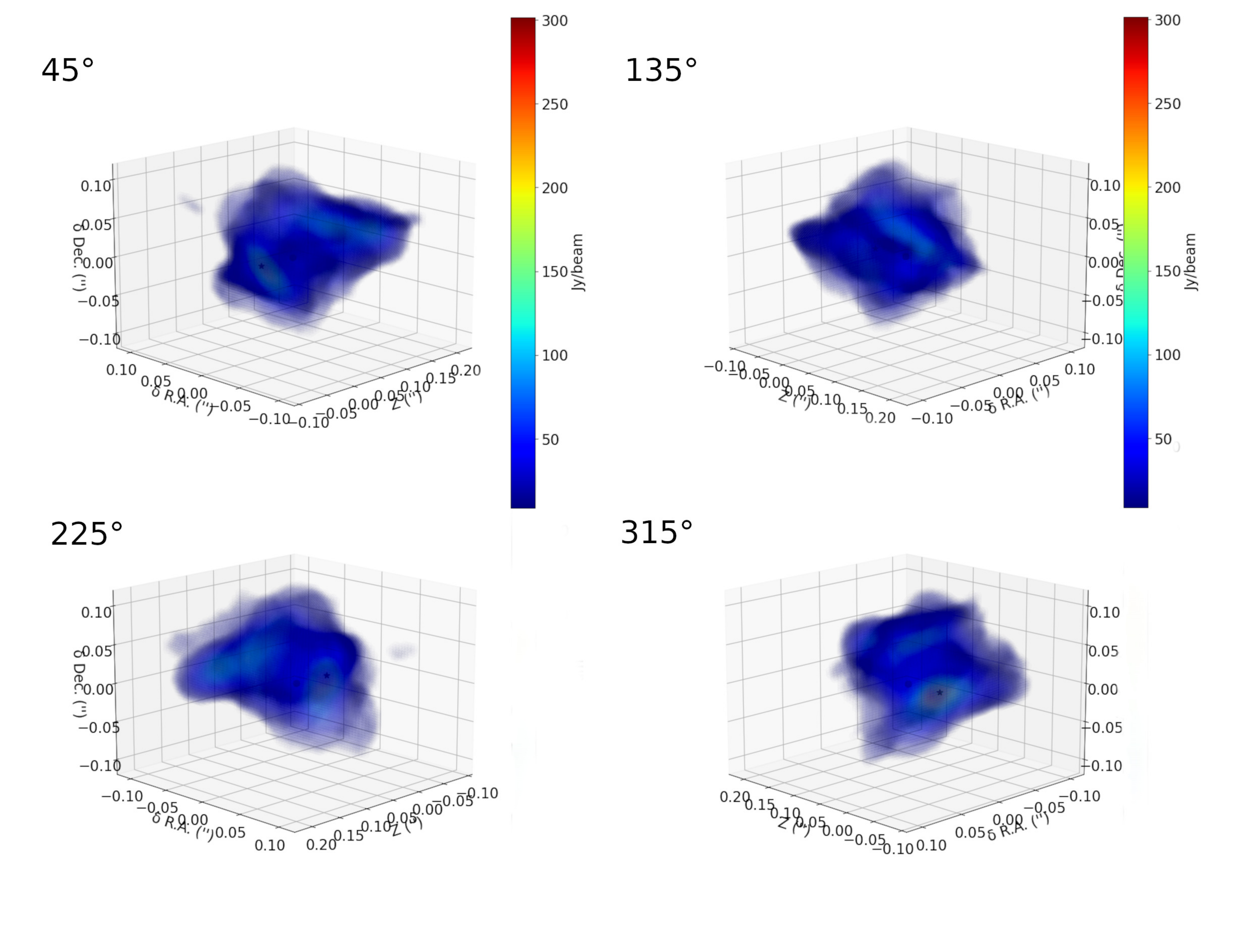}%H2S_H2Onewcenter.pdf}%NaCN_stack_azave.pdf} 

   \caption{Reconstruction of the ortho-H$_2$O $5_{5,0} - 6_{4,3}, v_2=1$ maser emission structure.
    For more detail, see the video presenting the rotation of the o-H$_2$O structure. The structure is created for the flux above $3\sigma$.}
              \label{water}%
\end{figure*}

At least two inner outflows have been identified from the o-H$_2$O $5_{5,0}-6_{4,3}$ $v_2=1$ emission. This 
emission is constrained to the innermost regions of the envelope around the star and is 
barely resolved (half power beam width, HPBW, $\sim 30 \times 20$mas).
The kinematic structure of this emission is particularly complex. In order to obtain 
a first approach to the structure, we assumed a  velocity field similar to that for the outer 
outflows. These outflows can be seen in Fig.\,\ref{water}. The resolution is 
not high enough to resolve the structure of the outflows.

\subsection{Position and systemic velocity of the star}

In order to identify, and confirm, the position-$V_\mathrm{LSR}$ of the central star we 
identified and traced the main axis of the different fast outflows. In the 3D videos labeled   ``lines''
we plotted the axis and the two positions for the central source as a filled star corresponding to a 
$V_\mathrm{LSR} = 12\kms$  and a filled circle for $V_\mathrm{LSR} = 22\kms$.
It can be seen that the former value seems more compatible with the ejections. 
This can be seen also in Figs.\,\ref{H2S-3d} and \ref{NaCl-3d}. 
The R.A. and Dec. for the central source are those obtained from the 
water maser, but it worth noting that assuming the R.A. and Dec. position 
derived from the continuum provides a similar result (i.e., $V_\mathrm{LSR} = 12\kms$).

\section{Origin of the ejecta}

\subsection{Regular ejecta}

By regular ejecta we mean that corresponding to the larger
part of circumstellar material (the standard mass ejection of the source) and
the ejecta that has  been ejected during most 
of the time and that is traced by species like SiO or SO.

SiO and SO emission appears both in the outer regions and in 
the innermost areas where H$_2$S is not present, suggesting that most 
of the time the mass ejection takes place in a steady way, as traced by 
these species, and that under certain conditions massive fast narrow 
ejections take place.

This ejecta suggest mass loss ejections at different positions angles 
but not clearly localized. A good candidate for these events would be 
the convective cells observed toward Betelgeuse \citep{Lim98} or the
recently proposed turbulence pressure \citep{Kee21}, which would form an 
extended atmosphere where dust can be formed. These ejections would 
create ubiquitous {(but not isotropic)} ejecta around the star. %The expansion velocity of this gas suggest that they are powered by radiation pressure on dust grains. 
In addition, these ejecta present a chemical footprint similar to 
that of high-mass O-rich AGB stars, in particular rich in SO and 
SO$_2$ lines. 
%The cavities and hollow-like structures observed in our interferometric maps of these lines can be related to fast outflows carving these ejecta.
These spectral maps present cavities and hollow structures  
that can be related to fast outflows, such as those traced by 
H$_2$S carving these ejecta.

As already mentioned, the velocity field of these ejecta is very complex, 
probably due to the presence of swept-up material  and the 
inhomogeneities of the mass loss ejection. However, the velocity 
field and the terminal velocity of the ejection are compatible with 
a mass ejection powered by radiation pressure on grains (i.e., 
$V_\infty = 30-40\kms$ for these type of objects). 
It has been found that other massive evolved stars present similar 
wide profiles that are related to the high luminosity of these 
objects \citep{gqtesis}. As mentioned above, the high-velocity wings 
arising from the outflows are not observed in the line profiles, due 
to projection effects, and it is the slowly expanding gas that is responsible 
for the line profile width observed.

\subsection{Sporadic outflows}

As shown above, more than six \emph{\emph{new}} outflows  have been detected 
with enough spatial and spectral resolution to reconstruct the structure of 
the ejecta and to derive their global characteristics. 
It has already been proposed that the mechanism driving these mass 
ejections might have a magnetic origin \citep{ogorman15,Vlemmings17}. 
Furthermore, the Hubble-like velocity field obtained suggests that the 
mechanism powering this ejection is not related to radiation pressure 
on dust grains formed on cold spots, but instead to energetic events 
similar to those responsible of the high-velocity outflows typical 
of post-AGB stars \citep{val2001}.

These jets are traced by H$_2$S and NaCl. The emission from these species has been associated 
with shocks and high-density regions, respectively \citep{Arce,NaCl_IRC}. 
Furthermore, H$_2$S has been claimed to be released from the  ice mantles 
of cold dust \citep{goico21}, where H$_2$S is formed and later on released 
into the gas due to shocks.

\begin{figure*}[th!]
   \centering
   \includegraphics[angle=0,width=18cm]{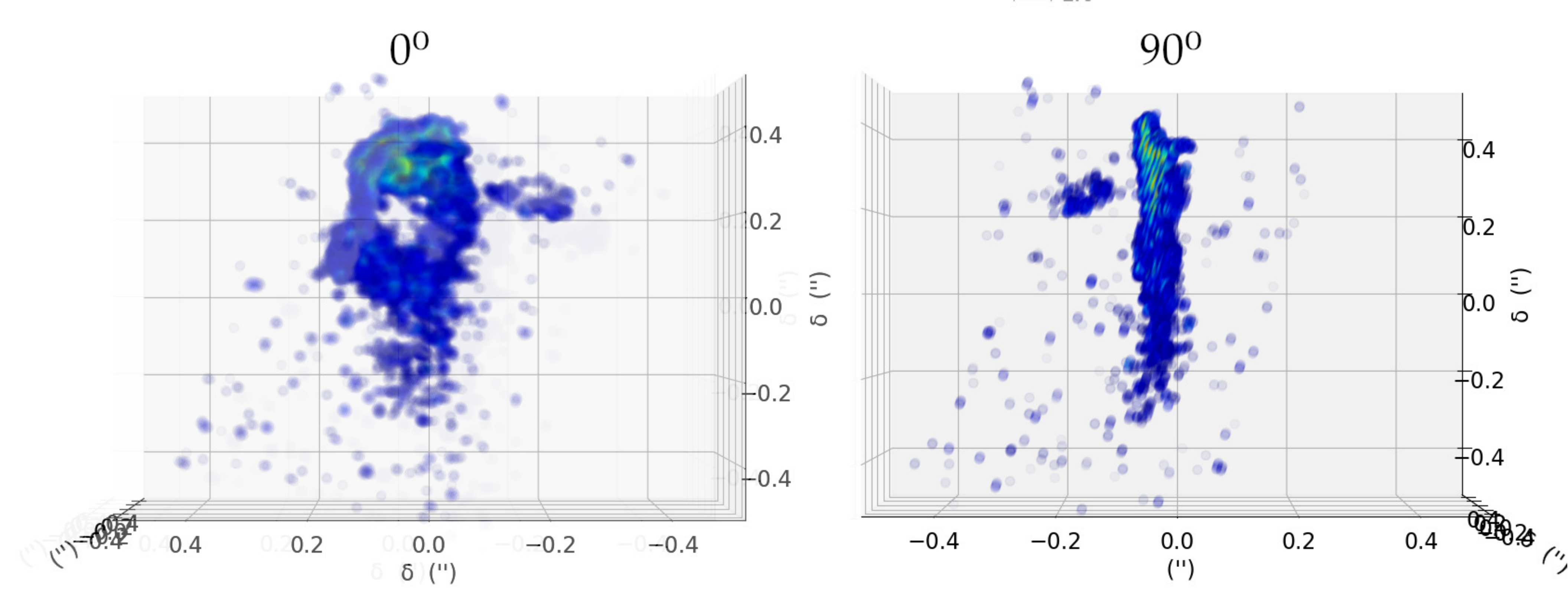}%NaCN_stack_azave.pdf} 
   \caption{Reconstruction of the H$_2$S emission presented face-on and tilted 90$^{\circ}$.}
              \label{C-structure}%
\end{figure*}

The extended outflows are well resolved and 
are found to have a flattened geometry (Fig. \ref{C-structure}) in one of the directions 
orthogonal to the axis of the ejection when compared with pPNe bipolar outflows 
\citep[see, e.g.,\ M1--92,][]{alcolea07}.
The ratio of width to height is $\sim\,6$.
A similar shape can be identified in the rest of the outflows observed (except for those
of H$_2$O which are not well resolved).
This flatness is not an effect of the velocity field assumed, and derived by the 
PV diagrams. It is reproduced for different outflows which expand along different 
directions. In addition, we have modified the velocity law from constant expansion 
to higher exponential orders, and this flatness is maintained. Therefore, we claim 
that the flatness seems to be an inherent characteristic of the fast outflows observed, and thus it has to 
be explored in order to understand the mass ejection of this object.

The SO structure that seems to surround the rear H$_2$S jet presents 
a shrinking tubular structure. This tubular structure also seems flatter 
in the vertical axis compared with the horizontal one 
(see Fig.\,\ref{SO_H2S}).
We inspected how certain mechanisms could be responsible for the formation of 
such flat structures.

\subsubsection{CME-like ejections. Linear momentum estimate.}

The presence of cold starspots alone could not generate the structures observed as 
they usually show a roughly circular shape,  
and the produced outflows are expected to have a %roughly 
 similar similar cross section orthogonal to their outflow axes. However, in the magnetically 
active regions of the stellar surface two mechanisms 
can generate outflows with the same shape as those observed 
in our ALMA maps.
These mechanisms are the sunspot groupings and the coronal mass ejections 
(CMEs), which create elongated features in the stellar surface.
The formation of a helmet streamer,
a close magnetic loop that generates bright loop-like structures,
and the subsequent CME in general takes place in 
active regions of the stellar surface where  cold-spot grouping also takes place. 
On the other hand, a cold-spot grouping is unlikely to generate the shape of the 
ejections  presented here, which in fact resembles the structure of the helmet 
streamers \citep{Glukhov_1997} and the final stages of the CMEs \citep{Gibson_1998}.
Moreover, the theoretical reconstruction of an observed fast solar CME \citep{jin13} %presented by e.g.,\ Jin et al. (2013)\cite{Jin} shows 
shows a structure more elongated in one of the dimensions %orthogonal to the axis of the outflow 
than in the other. In particular, this CME presents a structure
similar to that observed in NaCl emission (Fig.\,\ref{NaCl-3d}). 
It has been suggested that CME events might be an important source of mass loss in active
stars \citep{Argiroffi19}. 

Coronal mass ejections are expected to present cold plasma escaping from the stellar 
atmosphere \citep{Argiroffi19} and the presence of X-rays. However, X-rays have 
not been detected toward VY\,CMa \citep{MontezVYCMa}. This non-detection might be 
due to the disappearance of the CME eruption. As future collimated ejections 
or dimming events take place, target-of-opportunity X-ray observations 
might be useful to confirm the processes powering the ejections. 

%{\bf Linear momentum estimate.}
It is important to check, beyond the pure 
morphological similarity cited above, whether a CME can drive enough momentum to form 
a structure like those observed in VY\,CMa. We focus on the \emph{C} outflow since 
it is the most massive ejection \citep{ogorman15}, and especially because 
its properties can be accurately determined as it is isolated compared with the rest of the fast outflows. 

We adopt the higher value for the dust mass of \emph{C} derived by \citet{ogorman15} 
with a gas-to-dust ratio of 200. The total mass of the outflow is therefore 
0.05\ms. Other authors \citep{Kaminski19} have derived higher values for clump \emph{C}; 
however, these values are extremely uncertain.
The total momentum driven by this ejection assuming the Hubble velocity field 
deduced above and a constant mass distribution along the ejecta is
$P = 9.30 \times 10^{38}$\,g\,cm\,s$^{-1}$.

To estimate the momentum driven by a CME, we followed the discussion presented 
by \citet{Welsch18}. In this work the author estimates a momentum transfer of 1 
(g\,cm\,s$^{-1}$) per cm$^{2}$ per second for a magnetic field of 3\,G, 
being this momentum transfer rate proportional to the perpendicular magnetic 
field $B_\perp$.

The size of the cold spots in the RSGs has been found \citep{kunstler15} to reach sizes of 
$2.8 \times10^{23}$\,cm$^2$, and the time estimated for the 
formation of the \emph{C}-outflow is $\sim$50 years. The magnetic field 
needed to provide the amount of momentum derived above is $\sim$5kG. %5.45\,kG. 

The magnetic field of VY\,CMa has been estimated to reach a strength of 
1\,kG at the surface of the star \citep{Vlemmings02}, and more recently up to 
50\,kG \citep{Shinnaga17} assuming a toroidal magnetic configuration. 
These latter authors suggest that such strong magnetic fields are unlikely to occur
%on such object time 
since these objects are expected to have a slow rotation \citep{Monnier_1999}, a large radius, and low temperatures, 
and thus they suggest that the strength of the field should be somewhat lower and more locally enhanced.
However, higher rotation velocities have been observed in other RSGs \citep{Wheeler17} and planet engulfment
has been proposed as a source of increasing this velocity, causing an enhancement of the dynamo and
of the magnetic field \citep{Privitera16}. 
In addition, previous studies \citep{Vlemmings17} also suggested that magnetic 
field distribution seems to confirm that \emph{C} mass ejection is linked to the magnetic field. 

In summary, the structure, the magnetic field strength, and the position angle (P.A.) of the different jets, and also
the Hubble-like velocity field
 therefore seem compatible with outflows of magnetic origin,   for example coronal mass ejections.
A cold spot could also be responsible for the magnetically driven ejections, but not
as a consequence of  radiation pressure on newly formed dust, as mentioned above.

The formation of the fast outflows is estimated to have taken 
several tens of years, and thus stellar rotation might 
leave an imprint on the shape of the ejecta. The rotation velocity of 
RSGs is estimated  to be $\sim$1\kms\ \citep[typical for RSG stars,][]{Zhang12}, while for some objects 
such as Betelgeuse values as high as 15\kms\ have been derived \citep{Wheeler17}.
With a stellar radius of VY\,CMa of $R_{\star} = 9.9 \times 10^{13}$cm 
\citep{Wittkowski12} the effect of the rotation should be seen in a collimated 
outflow with the formation times derived. 
This scenario is explored in \ref{sprinklersec}.
 
\subsubsection{Sprinkler model}\label{sprinklersec}

If we assume that the flatness of the fast outflows is itself a 
consequence of a fast and collimated ejection combined with the 
stellar rotation, we should be able to find an angle that
would be compatible with all these ejections (i.e., the rotation 
axis). In this case, all the flat structures would be constrained 
within a single latitude angle ($\theta_i$) with respect to the rotation 
axis for each fast ejection.

\begin{figure*}[th!]
   \centering
   \hspace*{-3cm}\includegraphics[angle=0,width=23cm]{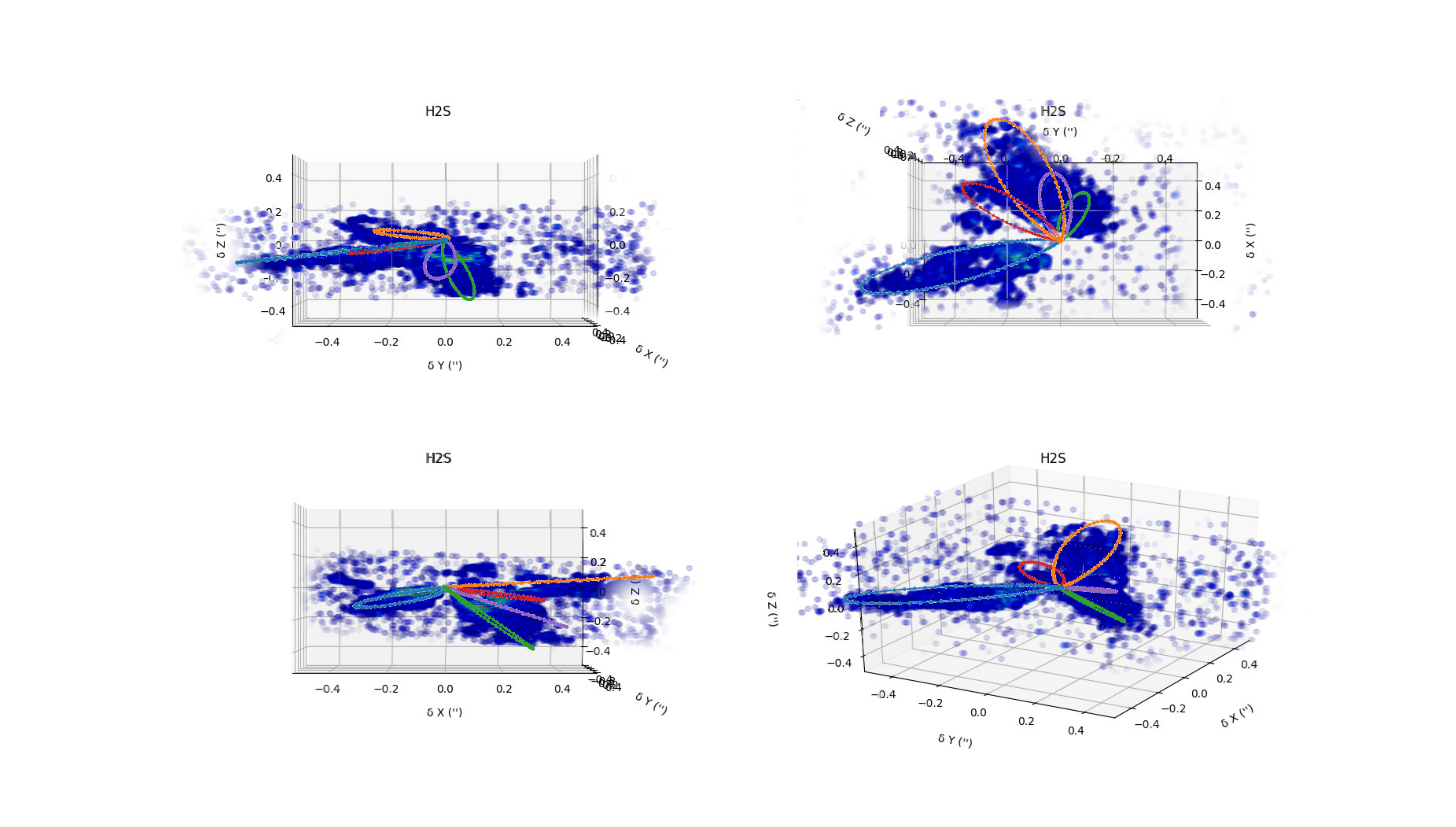} %esprikler.pdfe}%NaCN_stack_azave.pdf} 
   \caption{H$_2$S emission rotated so that the rotation axis corresponds 
   to the Z-axis, overplotted to the result of the sprinkler model applied to five ejections.}
              \label{sprinkler}%
\end{figure*}

In this sense, we have found a particular angle ($\sim$95$^\circ$) 
around the R.A. axis, which seems to be compatible with this scenario.
This means that the rotation axis of VY\, CMa would be approximately pole-on. 
 This rotated structure is presented in Fig.\,\ref{sprinkler}, with 
the rotation axis as $Z$.

In addition, we created a simple model to simulate the creation 
of these structures. The model consists of a position at the 
photosphere of the star from which a certain gas ejection takes 
place. We adopted the velocity law from the PV diagram, a rotation 
velocity of 1\,\kms\ \citep[typical for RSG stars;][]{Zhang12}, and adapted 
the lapse of activity of the ejection ($t_e$) with an expansion 
velocity moduled by a sinusoidal so it is at a minimum when the ejection 
starts and finishes (at $t_e$) and reaches a maximum at $t_e/2$.  
In any case, it is important to bear in mind that, as mentioned above, higher 
rotation velocities have been inferred for other objects, for example   $\sim$15\kms\ for 
Betelgeuse \citep{Wheeler17}.  

The parameters of each ejection are presented in Table 2. We initiate 
a rotation of the star with a given $v_{rot}$ at $\phi_0 =0$. The parameters 
$\theta_i$ and $\phi_i$ indicate the position and moment when the 
ejection starts, and $t_e$ the duration of the ejection; 
 $\theta_i$ is    the latitude of the ejecting spot and $\phi_i$
the rotation angle.
We first modeled the ejection corresponding to clump $C$ and used 
it as a starting point to fit the rest of the fast outflows. In particular 
we found that we had to reduce the velocity by 10\% for outflows 3--5,
although the ejection time could be maintained.
The result of our fitting of the sprinkler ejections to the observed 
H$_2$S structures is presented in Fig.\,\ref{sprinkler}.
The different extents of the ejections in Fig.\,\ref{sprinkler} are due to the fact that the ejections 
took place at different times.

\begin{table}
 \label{sprinktab}
 \begin{center}
 \caption{Parameters of the Sprinkler model outflows. The rotation 
velocity of the star is 1\kms. $V_{16}$ is the value for a Hubble-like 
expansion velocity at 10$^{16}$cm.% and its value is equivalent to that derived above at $0\secp1$ expect for outflow \#3.
 }
 \begin{tabular}{l c c c c}
 \hline\hline
  Outflow & $\theta_i$ ($^\circ$) & $\phi_i$ ($^\circ$) & $V_{16}$ (\kms) & $t_e$ (yr) \\
   \hline
  1             &       100     &       190     & 94.4  &       10              \\
  2             &       85      &       287     & 94.4  &       10              \\
  3             &       140     &       320     & 85.9  &       10              \\
  4             &       100     &       260     & 85.9  &       10              \\
  5             &       120     &       300     & 85.9  &       10              \\
\hline
\end{tabular}
\end{center}
\end{table}

This is a simple approximation that confirms that the molecular emission 
observed for the fast outflows can be formed in this manner. The ejection 
time depends directly on the rotation velocity. In general, we find that 
$t_e = 120/V_{rot}$[months\,s\,km$^{-1}$]. Therefore, for a $V_{rot}=15\kms$ 
the value of $t_e$ would be eight months. 
It is worth noting that the high magnetic field observed could be a consequence 
of an enhanced dynamo and due to high rotation velocities. In these cases we 
would expect low $t_e$.

We could also estimate the lapse between the different ejections. 
For a rotation velocity of 1\kms\ we found that the time between the ejections
was in the range $\sim$1 -- 6 years, being the time between the first and the last ejection
$\sim$11 years. For a $v_{rot}$ of 15\kms\ the range would be $\sim$1 -- 5 months and $\sim$9 
months between the first and last ejection.

It is also important to note that $t_e$ just takes into account the time when a particular
spot at the star is ejecting material. Once this ejection  stops, the structure   evolves given
an expansion velocity until  the observed structures are created. %In this particular simple model
%the total time until the structures reach the sizes observed around VY\,CMa is $\sim$ 175 years.

%It also suggest that the arcs observed at largest scales \cite{Humphreys07} 
%have the same origin as the fast outflows traced by H$_2$S.
It is worth noting that the arcs observed at larger scales \cite{Humphreys07} 
 resemble   the fast outflows traced by H$_2$S and may have been produced by the same mechanism in the past.

\section{Discussion}

The data we presented seem to support %prove 
the claim \citep{ogorman15} that the mass ejection in this object occurs 
in two different ways: a somewhat ubiquitous and steady mass ejection, 
with chemical properties of O-rich objects (SiO, SO, SO$_2$) and 
fast collimated outflows taking place at random directions in which 
the sulfur is mainly constrained into H$_2$S. The SO and SO$_2$ emission would 
trace swept-up material with high density.
% {\bf{VER MARKUYS WITTOSLI A VER}}
The former ejections are likely   due to convective cells, which 
would generate an irregular and extended atmosphere of circumstellar 
gas where most of the sulfur would be in the form of SO and SO$_2$, 
while another process, probably of magnetic origin, would generate 
fast outflows of gas and dust, which, when colliding with the 
surrounding material, would form H$_2$S by shocks or evaporate 
H$_2$S from dust \citep{Arce}. %This could also explain why only those O-rich stars with higher mass loss rates would present significant amounts of H$_2$S in their ejecta\cite{Danilovich}. 

Interestingly, \citet{Danilovich17} found that only  O-rich stars 
with higher mass loss rates  present significant amounts of H$_2$S in their ejecta. 
This, together with the above-mentioned relation between H$_2$S and jets, seems to suggest
that most massive stars will eventually present shocks that  alter their 
chemistry from the standard O-rich where most sulfur is in form of SO and SO$_2$ to 
a relevant presence of H$_2$S.

We propose that at a certain moment of its RSG evolution the surface 
of VY\,CMa became magnetically active and generated intense 
magnetically driven mass ejections. 
Certain processes have been proposed as being responsible for generating 
this type of activity in massive evolved stars, such as the 
appearance of a local dynamo in the giant convective cells \citep{auriene10} 
or an enhancement in the stellar dynamo by the engulfment of planets \citep{Privitera16}. 
These ejections would have generated anisotropical fast outflows like 
those presented in this work.

%+++++COMENTARIOS

Recent studies of other RSG stars reveal similar characteristics to VY\,CMa. In the
particular case of $\mu$\,Cep, CO observations also suggest two different mechanisms 
responsible for the ejecta around this object \citep{montarges19}, one responsible for the 
formation of a slow outflow and the other for the localized mass ejections. The authors 
of that work adopted a constant velocity field for both types of outflows, but they also 
suggest that this might not apply to $\mu$\,Cep clump \emph{C2}. This might also be the 
case of Betelgeuse \citep{levesque20}.

\section{Conclusions}

In summary, thanks to the spatial and spectral resolution of 
ALMA we were   able to detect previously unidentified spatial 
features of and found global characteristics of the outflows taking 
place in VY\,CMa. 
In particular the observations confirm that the structures 
observed are produced by two types of mass ejections:
\begin{itemize}
\item A slow wind, which might be powered 
by mechanisms such  as the large convective cells on the surface 
\citep{Lim98} that would form an extended warm atmosphere where dust 
formation and the subsequent radiation pressure on dust grains 
produce a smooth mass ejection. This ejection would not be either 
isotropic or strongly localized, but would generate an inhomogenous 
ejecta surrounding the object. The expansion velocity of this 
shell would be in the typical range for massive evolved stars 
(30--40\kms).
\item A collimated wind, with a Hubble expansion law and reaching 
velocities up to 100\kms\ similar to those observed in the post-AGB 
phase, which would carve the previous ejecta. These fast winds 
could not be due to radiation pressure on dust formed at long-lived 
cold spots, as indicated by the velocity field, but were caused by  other 
mechanisms. Coronal mass ejections or other magnetically driven 
mass ejection might be responsible for these ejections. These 
mass ejections could be related to local enhancements of the 
magnetic field \citep{Shinnaga17}.
\end{itemize}

In the general context of massive star evolution, and in 
particular of the mass loss processes determining the path 
of these objects during their last throes, the results  presented here
have a key relevance. 
The data presented show, for the first time, that mass 
loss within this phase can take place in two completely different 
ways. A future systematic study of RSGs and the characteristics of their ejecta, 
as well as detailed jet modeling to fit the structures will allow us 
to determine the characteristics favoring each type of mass event 
and the mechanism powering them. Two of the questions that should be addressed 
given the new mass loss paradigm suggested in this work     are whether  the slow expanding wind is the standard way these objects 
eject material, and if the  jets are restricted only to those
RSGs presenting higher rotation velocities and strong dynamos.

\begin{acknowledgements}
The research leading to these results has received funding from the European Research Council
under the European Union's Seventh Framework Programme (FP/2007-2013) / ERC Grant
Agreement n. 610256 (NANOCOSMOS). 
We would also like to thank the Spanish MINECO for funding support from grants CSD2009-00038, 
AYA2012-32032. M.A. also thanks for funding support from the 
Ram\'on y Cajal programme of Spanish MINECO (RyC-2014-16277). 
This publication is part of the ``I+D+i'' (research, development, and innovation) 
projects PID2019-107115GB-C21, and PID2019-106110GB-I00, and PID2020-117034RJ-I00, 
supported by the Spanish ``Ministerio de Ciencia e Innovaci\'on'' MCIN/AEI/10.13039/501100011033
\end{acknowledgements}

\bibliographystyle{aa} % style aa.bst
\bibliography{mybib}

\end{document}